\documentclass[aps,prd,onecolumn,groupedaddress,nofootinbib,amssymb]{revtex4}

\usepackage[latin1]{inputenc}
\usepackage{graphicx}
\usepackage[english]{babel}

\begin{document}

\title{ Gravitational waves from hyperbolic encounters}

\author{Salvatore Capozziello$^1$, Mariafelicia De Laurentis$^2$, Francesco de Paolis$^3$, G. Ingrosso$^3$,
Achille Nucita$^4$}

\affiliation{\it $^1$Dipartimento di Scienze fisiche, Università
di Napoli {}`` Federico II'', INFN Sez. di Napoli, Compl. Univ. di
Monte S. Angelo, Edificio G, Via Cinthia, I-80126, Napoli, Italy\\
$^2$Dipartimento di Fisica, Politecnico di Torino and INFN Sez. di
Torino, Corso Duca degli Abruzzi 24, I-10129 Torino, Italy\\
$^3$Dipartimento di Fisica di Università di Lecce and INFN Sez. di
Lecce, CP 193, I-73100 Lecce, Italy\\
$^4$XMM-Newton Science Operations Centre, ESAC, ESA, PO Box 78,
28691 Villanueva de la $\rm Ca\tilde{n}ada$, Madrid, Spain}

\begin{abstract}
The emission of gravitational waves from a system of massive
objects interacting on hyperbolic orbits is studied in the
quadrupole approximation. Analytic expressions are  derived for
the gravitational radiation luminosity, the total energy output
and the gravitational radiation amplitude. An estimation of the
expected number of events towards different targets (i.e. globular
clusters and the center of the Galaxy) is also given. In
particular, for a dense stellar cluster at the galactic center, a
rate up to one event per year is obtained.
\end{abstract}

%\pacs{95.85.Sz}
\maketitle


\vspace{4. mm}

General Relativity predicts that a system of interacting massive
objects emits gravitational waves which propagate in the vacuum
with the speed of light. A lot of studies have been devoted  to
the description of gravitational radiation emitted by  systems of
two interacting stars where  amplitude,  power and all physical
quantities of such a radiation strictly depend on the
configuration and the dynamics of the system. The seminal papers
by Peters and Mathews \cite{pm1,pm2}, which investigate the
gravitational wave emission by a binary system of stars (on
circular or elliptic orbits) in the quadrupole approximation, have
been extended in several directions (see e.g. \cite{schutz} and
references therein) in which the problem is studied by both
analytical and numerical approaches.

On the other hand, depending on their approaching energy,  stars
may interact also on unbound orbits (parabolic or hyperbolic) and,
in this case, one expects that gravitational waves are emitted
with a peculiar signature of  a "burst" wave-form having a maximum
in correspondence of the peri-astron distance. The problem of {\it
Bremsstrahlung}-like gravitational wave emission has been studied
in detail by Kovacs and Thorne \cite{kt} by considering stars
interacting on unbound orbits.

Here, we face this problem discussing the dynamics of such a
phenomenon in the  simplest case where stars interact on unbound
orbits and the peri-astron distance is much larger than the
Schwarzschild radius of the stars. In this case, the quadrupole
approximation holds and useful analytic quantities (as the energy
emitted by the system per unit time and gravitational wave
amplitude) can be derived.

In this letter, after reviewing the main features of hyperbolic
encounters between two stars, the emission of gravitational waves
is studied in the framework of the quadrupole approximation.
Particular attention is devoted to the gravitational radiation
luminosity. Then, we derive the expected gravitational wave-form
and discuss the detection of such events which could greatly
improve the statistics of possible gravitational wave sources.

\begin{figure}[ht]
\includegraphics[scale=0.6]{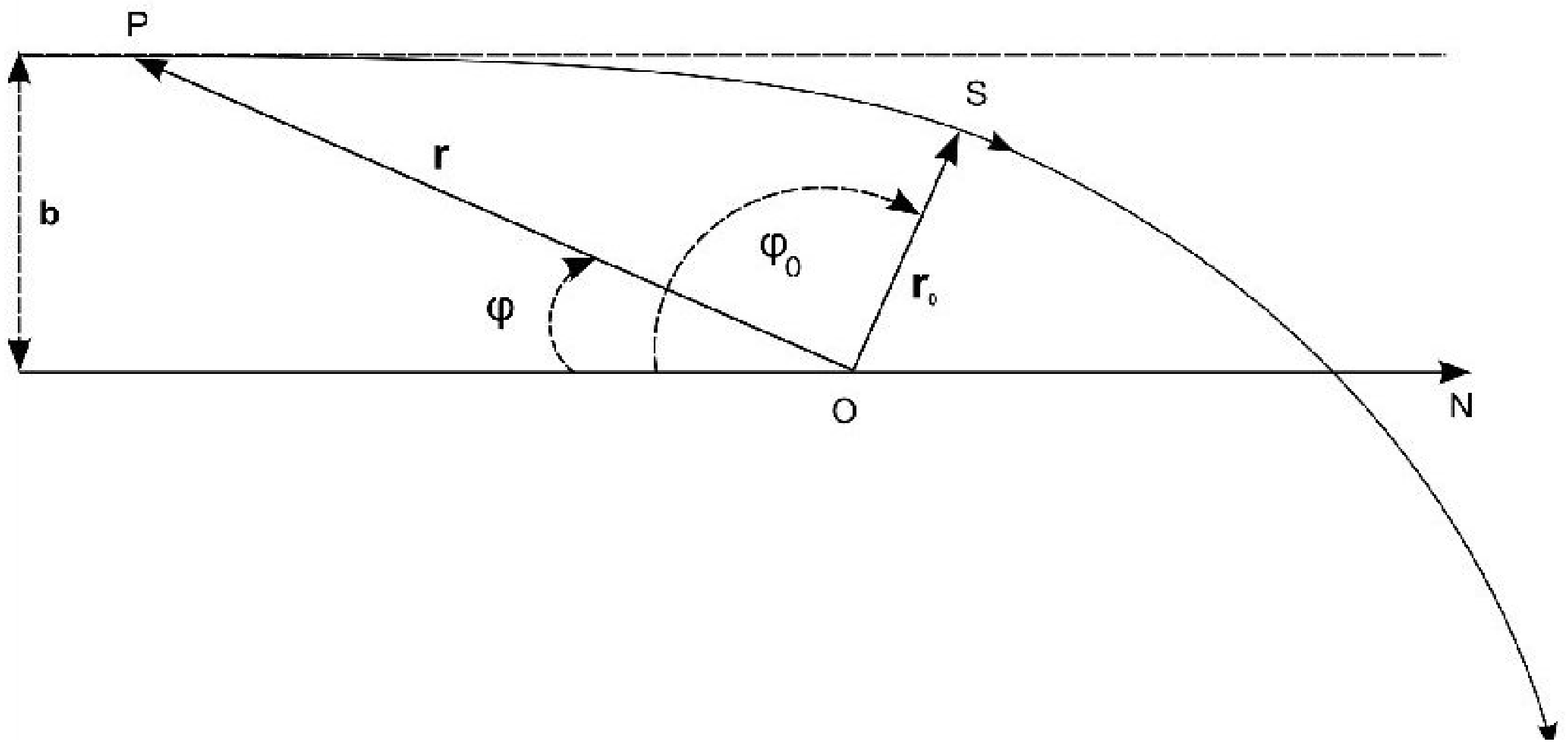}
\caption{The geometry of the hyperbolic encounter. The mass $M_1$
(in P) is moving on hyperbolic orbit (continuous line) with focus
in $O$ where the mass $M_2$ lies. The motion of $M_1$ is described
by the  vector radius $\textbf{r}$ and the polar angle $\phi$. The
vector radius $\textbf{r}_0$ (corresponding to the polar angle
$\phi _0$) represents the peri-astron distance, i.e. the distance
of the closest approach between the two interacting stars.}
\label{fig1}
\end{figure}

The study of  gravitational wave emission from massive objects
interacting on hyperbolic orbits can be started by analyzing the
geometry of  hyperbolic encounters. Without loss of generality,
let us consider a mass $M_1$ moving in the gravitational potential
$\Phi$ generated by a second mass $M_2$ at rest in the center $O$
of the  reference frame (see Fig. \ref{fig1}). Let $ON$ be a
reference direction in the orbital plane. Then, the position of
$M_1$ (in P) is specified by the vector radius  $\textbf{r}$ and
by the polar angle $\phi$ which the  vector radius forms with
$ON$, $\phi$ being measured in the direction of the star motion.
Obviously, both the  vector radius and the polar angle depend on
time as a consequence of the star motion, i.e.
$\textbf{r}=\textbf{r}(t)$ and $\phi=\phi(t)$.
%%%%%%%%%%%%%%%%%%%%%%%%%%%%%%%%%%%%%%%%%%%%%%%%%%%%
%%%%%%%%%%%%%%%%%%%%%%%%%%%%%%%%%%%%%%%%%%%%%%%%%%%
With this choice, the velocity $\textbf{v}$ of the mass $M_1$ can
be parameterized as
\begin{equation}
\textbf{v}=v_r\widehat{r}+v_{\phi}\widehat{\phi}~,
\end{equation}
where the radial and the tangent components of the velocity are,
respectively,
\begin{equation}
v_r=\frac{dr}{dt} ~~~~~~~~v_{\phi}=r \frac{d\phi}{dt}~.
\end{equation}
In this case, the total energy and the angular momentum, per unit
mass, read out
\begin{equation}
E=\frac{1}{2}v^2+\Phi (r) =
\frac{1}{2}\left(\frac{dr}{dt}\right)^2 +
\frac{1}{2}r^2\left(\frac{d\phi}{dt}\right)^2 +\Phi (r)
\label{eq:etot}
\end{equation}
and
\begin{equation}
L=r^2\frac{d\phi}{dt} \label{eq:momang1}~,
\end{equation}
respectively. At this point, it is useful to adopt the variable
$u=1/r$ so that eq. (\ref{eq:etot}) can be rewritten as
\begin{equation}
\left(\frac{du}{d\phi}\right)^2
+u^2+\frac{2\Phi(r)}{L^2}=\frac{2E}{L^2}= \mbox{const}.~,
\end{equation}
which, differentiated with respect to $u$, gives
\begin{equation}
\frac{d^2u}{d\phi^2}+u=- \frac{1}{L^2} \frac{d\Phi(u)}{du}~.
\end{equation}
The force acting on the unit mass is $F(r)=-\frac{GM_2}{r^2} =
-GM_2u^2$ which can be expressed as the gradient of the
gravitational potential $\Phi$. Hence, the last equation can be
recast in the form \cite{binney}
\begin{equation}
\frac{d^2u}{d\phi^2}+u= \frac{1}{L^2u^2} F \left(
\frac{1}{u}\right)~,\label{eq:diffe}
\end{equation}
and thus
\begin{equation}
\frac{d^2u}{d\phi^2}+u=\frac{GM_2}{L^2}~,
\end{equation}
whose general solution, giving the parameterized trajectory
followed by the unit mass particle, is
\begin{equation}
u(\phi)=C\cos (\phi-\phi_0) + \frac{GM_2}{L^2}~.
\label{eq:traiettoria1}
\end{equation}
Here, $C$ is a constant depending on the initial conditions and
$\phi_0$ is the polar angle corresponding to the peri-astron
distance, i.e. the distance of closest approach between the two
interacting particles. Note that after the interaction, the
particle is deflected by the angle $\phi_d=2\phi_0$.

For a system of two objects (of mass $M_1$ and $M_2$) interacting
on hyperbolic orbits, the two-body problem can be reduced to the
problem of a reduced mass particle $\mu =M_1M_2/(M_1+M_2)$ moving
in the gravitational field generated by a total fictitious mass
$M=M_1+M_2$. Let us assume that $M_2$ is at rest while $M_1$ is
moving with initial (at  infinite distance) velocity $v_0$ with
impact parameter $b$ (see Fig. \ref{fig1}).

Considering what previously stated, it is then clear that the
trajectory followed by the reduced mass particle is
\begin{equation}
u(\phi)=C\cos(\phi-\phi_0)+\frac{G(M_1+M_2)}{L^2}~.
\label{eq:traiettoria2}
\end{equation}
Differentiating the previous relation with respect to the time $t$
and turning back to the variable $r$, one obtains
\begin{equation}
\frac{dr}{dt}=C L\sin(\phi-\phi_0)~, \label{drdt}
\end{equation}
by which it possible to determine the value of the constant $C$
through the initial conditions of the motion, i.e.
\begin{equation}
C = \frac {v_0} {L \sin (\phi_0)}~,
\end{equation}
which can be written, being $L=bv_0$, as
\begin{equation}
C= \frac{1} {b\sin {\phi_0}}~.\label{eq:c1}
\end{equation}

Alternatively, the initial conditions of the motion can be used
directly in eq. (\ref{eq:traiettoria2}). In this case, one finds
\begin{equation}
C=-\frac{G(M_1+M_2)} {b^2\cos{\phi_0}}~, \label{eq:c2}
\end{equation}
which compared with eq. (\ref{eq:c1}) gives the  result
\begin{equation}
\tan{\phi_0}=-\frac{bv_0^2}{G(M_1+M_2)}~. \label{tanphi0}
\end{equation}
Clearly, eq.(\ref{eq:c1}) is meaningless for $\phi_0=90^o$, which
means that, for this critical value, there is no interaction
between the stars which are at an infinite distance each other. In
other words, stars are very far when $\phi$ approaches to $90^o$.

Finally, the equation of the orbit followed by the reduced mass
particle turns out to be \cite{smart}
\begin{equation}
r = \frac{b\sin{\phi_0}}{\cos(\phi-\phi_0)-\cos{\phi_0}}~,
\label{eq:traiettoria}
\end{equation}
which allows to determine the modulus of the  radius vector
$\textbf{r}$ as a function of the polar angle $\phi$ once the
initial conditions are known. With this set of equations at hand,
we can estimate both the gravitational wave luminosity of the
system and the gravitational radiation wave-form in the quadrupole
approximation. We are adopting such an approximation since we are
considering "binary" systems whose distance is larger than the
"capture" distance. In other words, we are considering situations
and initial conditions where pointlike approximation of stars
holds and the system remains unbounded.  In such cases, quadrupole
approximation works and results are reasonable \cite{blanchet}.

The Einstein field equations give a description of how the
curvature of space-time, at any event, is related to the
energy-momentum distribution at that event. In the weak field
approximation, it is  found that systems of massive moving objects
produce gravitational waves which propagate in the vacuum with the
speed of light. It can be shown that the energy emitted per unit
time, in the form of gravitational radiation (after integrating on
all the gravitational wave polarization states), is \cite{landau}

\begin{equation}
\frac{dE}{dt}=-\frac{G\left\langle D_{ij}^{(3)}D^{(3)ij}\right\rangle }{45c^{5}}\label{eq:dEdt}\end{equation}

where the dot represents the differentiation with respect to time,
the symbol  $\langle \rangle$ indicates the scalar product and the
quadrupole mass tensor $D_{ij}$ is defined as
\begin{equation}
D_{ij}=\sum _a
m_a(3x_a^ix_a^j-\delta_{ij}r_a^2)~,\label{qmasstensor}
\end{equation}
$r_a$ being the modulus of the vector radius of the $a-th$
particle.

It is then possible to estimate the amount of energy emitted in
the form of gravitational waves from a system of massive objects
interacting on hyperbolic orbits. In this case, the components of
the quadrupole mass tensor are
\begin{equation}
\begin{array}{llll}
D_{11}=\mu r^2(3\cos{^2\phi}-1)~,\\ \\
D_{22}=\mu r^2(3\sin{^2\phi}-1)~,\\ \\
D_{12}=D_{21}=3\mu r^2 \cos\phi \sin\phi~, \\ \\
D_{33}=-\mu r^2,
\end{array}
\end{equation}
which can be differentiated with respect to time as required in
eq. (\ref{eq:dEdt}). In doing this, we can use some useful
relations derived above, in particular eqs. (\ref{eq:momang1}),
(\ref{drdt}), (\ref{tanphi0}) and (\ref{eq:traiettoria}). It is
straightforward to show that
\begin{equation}
 D_{ij}^{(3)}D^{(3)ij}=\frac{32L^6\mu^2}{b^8}f(\phi,\phi_0)~,
\label{eq:tensore}
\end{equation}
where $f(\phi,\phi_0)$ is given by
\begin{equation}
\begin{array}{lll}
\displaystyle{f(\phi,\phi_0)=\displaystyle{\sin^4 (\phi_0- \phi/2)
\sin^4 (\phi/2)}~\tan^{-2} \phi_0~\sin^{-6} \phi_0 }\\
\displaystyle{~~~~~~~~~~~~~~~\times \left[150+72\cos{2\phi_0}+ 66\cos{2(\phi_0-\phi)} \right.} \\
\displaystyle{~~~~~~~~~~~~~~~\left.-144\cos{(2\phi_0-\phi)}-
144\cos{\phi} \right]}~.
\end{array}
\label{eq:funzione}
\end{equation}
Hence, the energy emitted by the system per unit time is
\begin{equation}
\frac{dE}{dt}=-\frac{32GL^6\mu^2}{45c^5b^8}f(\phi,\phi_0)~,
\label{eq:energiaiperbole1}
\end{equation}
or, equivalently,
\begin{equation}
\frac{dE}{dt}=-\frac{32Gv_0^6\mu^2}{45c^5b^2}f(\phi,\phi_0)~,
\label{eq:energiaiperbole2}
\end{equation}
which, for $M_1=M_2$, can be rewritten as
\begin{equation}
\frac{dE}{dt}=-\frac{4r_sv_0^6m}{45c^3b^2}f(\phi,\phi_0)~,
\label{eq:energiaiperbole3}
\end{equation}
$r_s$ being the Schwarzschild radius of the mass.

The total energy emitted in the form of gravitational radiation
during the interaction is given by
\begin{equation}
\Delta E=\int^{\infty}_0 |\frac{dE}{dt}| dt~.
\end{equation}
Since eq. (\ref{eq:momang1}) holds, we can adopt the angle $\phi$
as a convenient integration variable. In this case, the energy
emitted for $\phi_1<\phi<\phi_2$ is
\begin{equation}
\Delta E(\phi_1,\phi_2)
=\frac{4r_smv_0^5}{45c^3b}\int^{\phi_2}_{\phi_1}
\frac{\sin{^2\phi_0}~f(\phi,\phi_0)}{[\cos{(\phi-\phi_0)}-\cos{\phi_0}]^2}
~d\phi~,\label{eq:integraleenergia}
\end{equation}
and the total energy can be determined from the previous relation
in the limits $\phi_1\rightarrow 0$ and $\phi_2\rightarrow2
\phi_0$. Thus, one has
\begin{equation}
\Delta E=\frac{v_0^5r_sm}{c^3}F(b,v_0)~,
\end{equation}
where $F(b,v_0)$ only depends on the initial conditions  and it is
given by
\begin{equation}
\begin{array}{ll}
\displaystyle{F(b,v_0)=[720b \tan^2{\phi _0} \sin ^4{\phi _0}]^{-1}\times}\left(2628 \phi_0\right.\\
~~~~~~~~\left. +2328 \phi_0 \cos{2\phi_0} +
    144 \phi_0 \cos{4 \phi_0}\right. \\
~~~~~~~~\left. - 1948 \sin{2 \phi_0} -
    301 \sin{4 \phi_0}\right)~.
\end{array}\end{equation}
In other words, the gravitational radiation luminosity strictly
depends on the configuration and kinematics of the binary system
and it improves at short $b$ and high $v_0$.

Direct signatures of gravitational radiation are its amplitude and
its wave-form. In other words, the identification of a
gravitational radiation signal is strictly related to the accurate
selection of  the shape of wave-forms by interferometers or any
possible detection tool. Such an achievement could give
information on the nature of the gravitational wave source, on the
propagating medium, and even, in general,  on the gravitational
theory producing such a radiation \cite{cap,ccd}. It is well known
that the amplitude of the gravitational waves can be evaluated as
\begin{equation}
h^{jk}(t,R)=\frac{2G}{Rc^4}\ddot{D}^{jk}~, \label{ampli1}
\end{equation}
$R$ being the distance between the source and the observer and
$\{j,k\}=1,2$.

Considering our binary system and the single components of
eq.(\ref{ampli1}), it is straightforward to show that
\begin{equation}
\begin{array}{llllllll}
h^{11}=\frac{2G}{Rc^4}\frac{\csc ^2(\phi _0)\mu
v_0^2}{4}\left[13\cos\phi-12\cos 2\phi +3\cos3\phi\right.\\
\left.-2\cos(\phi-2\phi _0)+3\cos (3\phi-2\phi_0)-12\cos(2
\phi_0)\right.\\ \left.-6\cos 2(\phi
_0-\phi)-6\cos2(\phi+\phi_0)+15 \cos (\phi +2 \phi _0)+4\right]~,
\\ \\
h^{22}=\frac{2G}{Rc^4} \frac{\csc ^2(\phi _0)\mu v_0^2}{4}
\left[-17 \cos\phi+12 \cos 2\phi-3 \cos 3 \phi\right.\\
\left.-2 \cos (\phi -2 \phi _0)-3 \cos (3 \phi -2 \phi _0)+12 \cos
2 \phi _0\right.\\
\left.+6 \cos 2 (\phi _0-\phi)+6 \cos 2(\phi +\phi _0)-15
\cos(\phi +2 \phi _0)+4\right]~,
\\ \\
h^{12}=h^{21}=\frac{2G}{Rc^4} 3\mu v_0^2 \csc ^2\phi _0 \sin
^2\phi/2 \left[2 \sin \phi-\sin 2
\phi \right. \\
\left. -\sin 2 \phi _0+\sin 2(\phi _0-\phi)+2 \sin(\phi +2
\phi_0)\right]~,\label{param}
\end{array}
\end{equation}
so that the expected strain amplitude
$h\simeq(h_{11}^2+h_{22}^2+2h_{12}^2)^{1/2}$ turns out to be

\begin{eqnarray}
h=\frac{2G}{Rc^4} \mu v_0^2 \csc ^2\phi _0 \left\{2 \left[59
\cos2(\phi _0-\phi)-\cos\phi(54 \cos \left(2 \phi
_0\right)+101)\right] \cos ^2\phi _0\right.\nonumber\\
\left.-9 \cos \left(3 \phi -4 \phi_0\right)-9 \cos
\left(3 \phi -2 \phi_0 \right)+95 \cos2\phi_0 +9 \cos4\phi_0- \right.\nonumber\\
\left. \sin\phi\left[101\sin 2\phi_0 +27 \sin
4\phi_0\right]+106\right\}^{1/2}~,
\end{eqnarray}
which, as before, strictly depends on the initial conditions of
the stellar encounter. A remark is in order at this point. A
monochromatic gravitational wave has, at most, two independent
degrees of freedom while eq. (\ref{param}) seems  to show three
independent parameters associated to the amplitude of the
corresponding wave. This is not true since, as usual in the TT
gauge, we have $h_+$ and $h_{\times}$ being $h_+ = h_{11}+h_{22}$
and $h_{\times} = h_{12}+h_{21}$. For details, see
\cite{blanchet}.

As an example, the amplitude of gravitational wave is sketched in
Fig.2 for a stellar encounter close to the Galactic Center. The
adopted initial parameters are typical of a close hyperbolic
impact and are assumed to be $b=1$ AU  and $v_{0}=200$ Km$s^{-1}$,
respectively. Here,  we have fixed $M_{1}=M_{2}=1.4M_{\odot}$.

\begin{figure}[ht]
\includegraphics[scale=0.6]{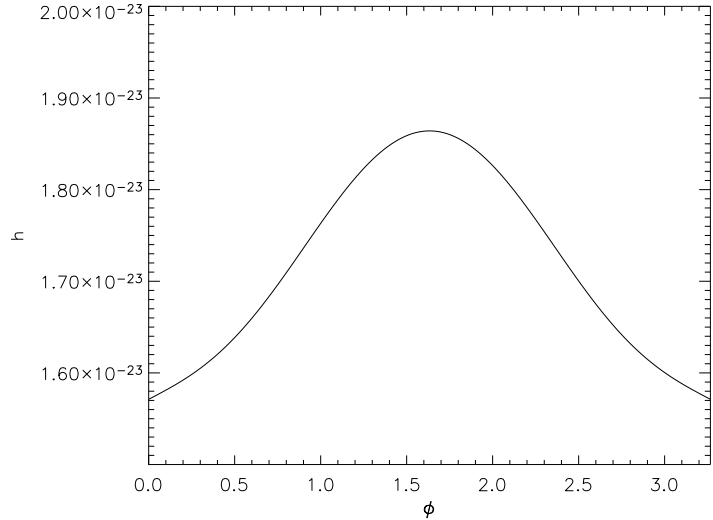}
\caption{The gravitational wave-form is sketched as a function of
the polar angle $\phi$ for some values of both the impact
parameter and velocity. In particular, we have fixed
$M_{1}=M_{2}=1.4M_{\odot}$. $M_{2}$ is considered at rest while
$M_{1}$ is moving with initial velocity $v_{0}=200$ Km$s^{-1}$ and
an impact parameter $b=1$ AU. The distance of the source has been
assumed to be $r=8$ kpc. As expected, the wave-form has a maximum
in correspondence of the peri-astron distance. } \label{fig:og}
\end{figure}

In the following, we give an estimate of the rate of stellar
encounters on hyperbolic orbits in some interesting astrophysical
conditions as a typical globular cluster or towards the bulge of
our Galaxy.

Let us consider a generic star cluster with mass density profile
$\rho(r)$ composed, for the sake of simplicity, by $N_*$ stars of
equal mass $m$ moving with relative velocities $v_{\rm rel}(r)$.
Then, the event rate $\Gamma$, that is the number of interacting
stars on hyperbolic orbits per unit time, is given by
\begin{equation}
\Gamma=\int_{R_{\rm
0}}^{R}\left[\frac{\rho(r)}{m}\right]^2\sigma(r)v_{\rm rel}(r)4\pi
r^2 dr~.
\end{equation}
Here, $\rho (r)$ is the cluster density profile (which we assume
to follow a Plummer model \cite{binney}), $R$ its radius and
$\sigma(r)\simeq \pi [b_1(r)^2-b_2(r)^2]$ the typical cross
section of the hyperbolic encounter at distance $r$ from the
cluster center (see \cite{clayton} for details). In evaluating the
rate, we are considering only those hyperbolic encounters
producing gravitational waves, for example,  in the LISA range,
i.e. between $10^{-4}$ and $10^{-2}$ Hz (see e.g. \cite{rubbo}).
The integral on the right hand of the previous equation can be
approximately solved and put in the form
\begin{equation}
\Gamma \simeq  5.5\times 10^{-10} \left(\frac{M}{10^5 {\rm
M_{\odot}}}\right)^2 \left(\frac{v}{10 {\rm km s^{-1}}}\right)
\left(\frac{\sigma}{UA^2}\right) \left(\frac{{\rm 10
pc}}{R}\right)^3 {\rm yrs^{-1}}.
\end{equation}
For a typical globular cluster (GC) around the Galactic Center,
the expected event rate is of the order of $2\times 10^{-9}$
yrs$^{-1}$ which may be increased at least by a factor $\simeq
100$ if one considers the number of GCs in the Galaxy. If the
stellar cluster at the galactic center is taken into account and
assuming $M\simeq 3\times 10^6$ M$_{\odot}$, $v\simeq $ 150 km
s$^{-1}$ and $R\simeq$ 10 pc, one expects to have $\simeq 10^{-5}$
open orbit encounters per year. If a cluster with total mass
$M\simeq 10^6$ M$_{\odot}$, $v\simeq $ 150 km s$^{-1}$ and
$R\simeq$ 0.1 pc is considered, an event rate number of the order
of unity per year is obtained. These values could be realistically
achieved by data coming from the forthcoming  space interferometer
LISA.

In this letter, the gravitational wave emission on hyperbolic
stellar encounters has been analyzed. In particular, we have taken
into account the expected luminosity and the  strain amplitude of
gravitational radiation produced in a tight hyperbolic impact
where two massive objects of $1.4M_{\odot}$ closely interact at an
impact distance of $1AU$. Due to the high probability of such
encounters inside rich stellar fields (e.g. globular clusters,
bulges of galaxies and so on), the presented approach could highly
contribute to enlarge the classes of gravitational wave sources
(in particular, of dynamical phenomena capable of producing
gravitational waves). In particular, a detailed theory of stellar
orbits could improve the statistics of possible gravitational wave
sources.

\end{document}